\RequirePackage{fix-cm}
\documentclass[smallextended]{svjour3}       
\smartqed  
\usepackage{graphicx}
\journalname{Journal of Low Temperature Physics}

\begin{document}

\title{Magnetic signatures of quantum critical points of the ferrimagnetic mixed spin-(1/2, $S$) Heisenberg chains at finite temperatures\thanks{This work was financially supported by the grant of The Ministry of Education, Science, Research and Sport of the Slovak Republic under the contract No. VEGA 1/0043/16, as well as, by grants of the Slovak Research and Development Agency provided under Contract Nos. APVV-0097-12 and APVV-14-0073.}}
\titlerunning{Magnetic signatures of quantum critical points of the Heisenberg chains}   
\author{Jozef Stre\v{c}ka \and Taras Verkholyak}
\institute{Jozef Stre\v{c}ka \at
           Institute of Physics, Faculty of Science of P. J. \v{S}af\'arik University, \\ 
           Park Angelinum 9, 040 01 Ko\v{s}ice, Slovak Republic \\                         
           \email{jozef.strecka@upjs.sk}           
           \and
           Taras Verkholyak \at
           Institute for Condensed Matter Physics, NASU, \\
           1 Svientsitskii Street, 790 11 L'viv-11, Ukraine \\
           \email{werch@icmp.lviv.ua}}

\date{Received: date / Accepted: date}

\maketitle

\begin{abstract}
Magnetic properties of the ferrimagnetic mixed spin-(1/2,$S$) Heisenberg chains are examined 
using quantum Monte Carlo simulations for two different quantum spin numbers $S=1$ and 3/2. The calculated magnetization curves at finite temperatures are confronted with zero-temperature magnetization data obtained within density-matrix renormalization group method, which imply an existence of two quantum critical points determining a breakdown of the gapped Lieb-Mattis ferrimagnetic phase and Tomonaga-Luttinger spin-liquid phase, respectively. While a square-root behavior of the magnetization accompanying each quantum critical point is gradually smoothed upon rising temperature, the susceptibility and isothermal entropy change data provide a stronger evidence of the quantum critical points at finite temperatures through marked local maxima and minima, respectively. 
\end{abstract}

\section{Introduction}
\label{intro}
Quantum phase transitions traditionally attract a great deal of attention, because they are often accompanied with several remarkable signatures experimentally accessible at nonzero temperatures \cite{sac99}. One-dimensional quantum spin chains provide notable examples of condensed matter systems, which bring a deeper understanding into exotic forms of magnetism nearby quantum critical points based on rigorous calculations \cite{mat93}. Apart from conventional solid-state compounds affording experimental realizations of the quantum spin chains \cite{mil01}, one may simulate a quantum phase transition in spin chains through ultracold atoms confined in an optical lattice \cite{sim11}. Fractional magnetization plateaux \cite{hon04}, magnetization cusp singularities \cite{oku02} and spin-liquid ground states \cite{hon04,oku02,mis08,zho16} can be regarded as the most profound manifestations of zero-temperature magnetization curves of the quantum spin chains.

The ferrimagnetic mixed spin-($s$, $S$) Heisenberg  chains \cite{kur98,yam99,sak99,iva00,hon00,yam00,sak02,ten11,iva16,str16} with regularly alternating spins $s=1/2$ and $S>1/2$ display the intermediate magnetization plateau inherent to the gapped Lieb-Mattis ferrimagnetic ground state, as well as, the gapless Tomonaga-Luttinger spin-liquid phase. It has been argued \cite{str16} on the grounds of Lieb-Mattis theorem \cite{lie62} and Oshikawa-Yamanaka-Affleck rule \cite{oya97} that the ferrimagnetic mixed spin-(1/2,$S$) Heisenberg chains should exhibit just one plateau at the following value of the total magnetization $m/m_s = (2S-1)/(2S+1)$ normalized with respect to its saturation. The gapped Lieb-Mattis ferrimagnetic ground state breaks down at a quantum phase transition towards the gapless Tomonaga-Luttinger spin-liquid phase, which is accompanied with a singular square-root behavior of the magnetization. The same type of the magnetization singularity also appears just below the saturation field, which determines a quantum phase transition between the Tomonaga-Luttinger spin liquid and the fully polarized phase. In the present work we will examine magnetic signatures of the quantum critical points of the ferrimagnetic mixed spin-(1/2,$S$) Heisenberg chains at finite temperatures by making use of quantum Monte Carlo simulations.

\section{Model and method}
\label{model}

Let us consider the ferrimagnetic mixed spin-(1/2,$S$) Heisenberg chains defined through the Hamiltonian
\begin{eqnarray}
\hat{\cal H} = J \sum_{j=1}^L \hat{\bf S}_j \cdot (\hat{\bf s}_j + \hat{\bf s}_{j+1}) - h \sum_{j=1}^L (\hat{S}_j^z + \hat{s}_j^z),
\label{ham}
\end{eqnarray}
where $\hat{\bf s}_j \equiv (\hat{s}_j^x,\hat{s}_j^y,\hat{s}_j^z)$ and $\hat{\bf S}_j \equiv (\hat{S}_j^x,\hat{S}_j^y,\hat{S}_j^z)$ denote the standard spin-1/2 and spin-$S$ operators, respectively. The first term in the Hamiltonian (\ref{ham}) takes into account the antiferromagnetic Heisenberg interaction $J>0$ between the nearest-neighbor spins and the second term $h = g \mu_{\rm B} H$ incorporating the equal Land\'e g-factors $g_s = g_S = g$ and Bohr magneton $\mu_{\rm B}$ accounts for the Zeemann's energy of individual magnetic moments in an external magnetic field. Since the elementary unit contains two spins an overall chain length is $2L$, whereas a translational invariance is ensured 
by the choice of periodic boundary conditions $\hat{\bf s}_{L+1} \equiv \hat{\bf s}_1$. 

To explore magnetic properties of the ferrimagnetic mixed spin-(1/2,$S$) Heisenberg chains at nonzero temperatures, we have implemented a directed loop algorithm in the stochastic series expansion representation of the quantum Monte Carlo (QMC) method \cite{san99} from Algorithms and Libraries for Physics Simulations (ALPS) project \cite{bau11}. The QMC method allows a straightforward calculation of the magnetization data at finite temperatures, which will be also confronted with recent zero-temperature magnetization data calculated within the density-matrix renormalization group (DMRG) method \cite{str16} serving as a useful benchmark at low enough temperatures. The susceptibility of the ferrimagnetic mixed spin-(1/2,$S$) Heisenberg chains can also be directly calculated from a directed loop algorithm of QMC method, while the isothermal entropy change can be obtained from the relevant magnetization data using the Maxwell's relation
\begin{eqnarray}
\Delta S_T = \int_{0}^{h} \left(\frac{\partial m}{\partial T}\right)_{h} {\rm d} h.
\label{iso}
\end{eqnarray}
To avoid an extrapolation due to finite-size effects, we have performed QMC simulations for a sufficiently large system size with up to $L=128$ units (256 spins), whereas adequate numerical accuracy was achieved through 750 000 Monte Carlo steps used in addition to 
150 000 steps for thermalization.
 
\section{Results and discussion}
\label{result}

\begin{figure}
\includegraphics[width=0.6\textwidth]{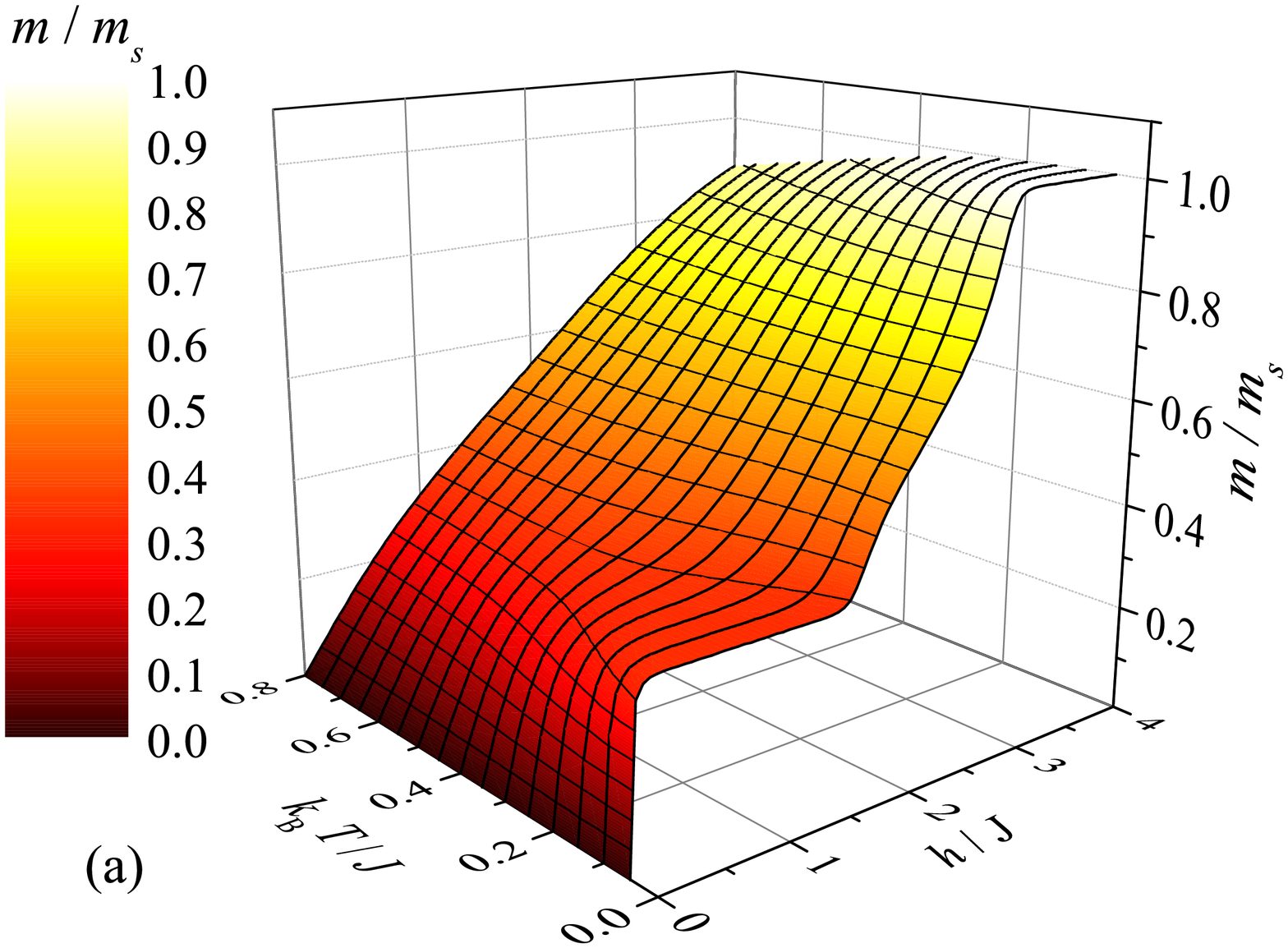}
\hspace{-1.7cm}
\includegraphics[width=0.6\textwidth]{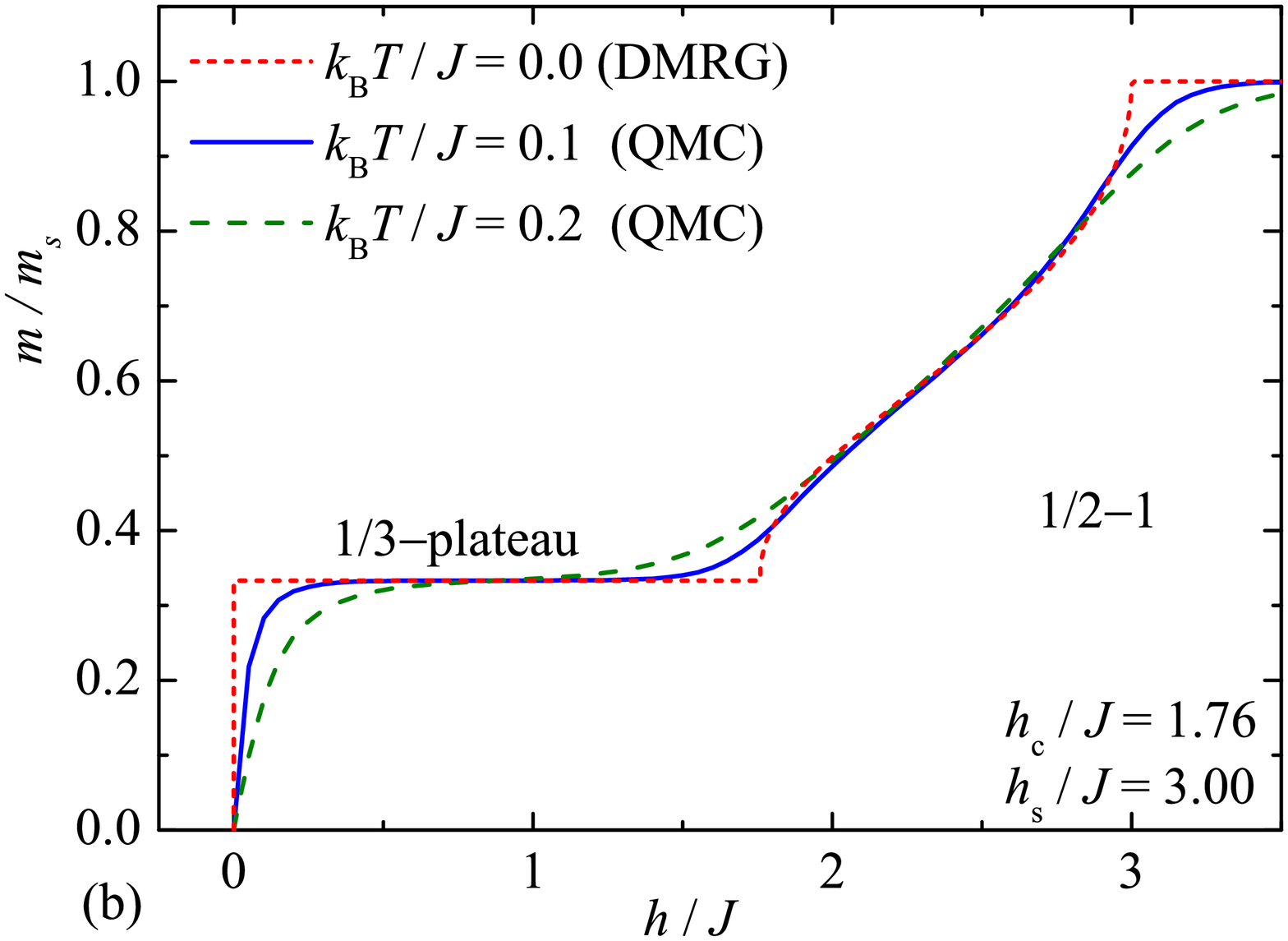}
\vspace*{-0.8cm}
\caption{The total magnetization of the ferrimagnetic mixed spin-(1/2,1) Heisenberg chain normalized with respect to its saturation value: (a) 3D surface plot as a function of temperature and magnetic field constructed from QMC data; (b) Zero-temperature DMRG data versus QMC magnetization curves at low enough temperatures.}
\label{fig1}       
\end{figure}

Let us proceed to a discussion of the most interesting results for the magnetization and susceptibility data of the ferrimagnetic mixed spin-(1/2,$S$) Heisenberg chains. Fig. \ref{fig1}(a) shows a three-dimensional (3D) plot of the magnetization of the ferrimagnetic mixed spin-(1/2,1) Heisenberg chain against the magnetic field and temperature. As one can see, the one-third plateau due to the gapped Lieb-Mattis ferrimagnetic ground state diminishes upon increasing of temperature until it becomes completely indiscernible above certain temperature $k_{\rm B} T/J \approx 0.5$. To explore the temperature effect in more detail, Fig. \ref{fig1}(b) compares a zero-temperature magnetization curve obtained within DMRG method \cite{str16} with low-temperature magnetization data stemming from QMC simulations. It is quite evident that the square-root singularity of the magnetization, which emerges at both quantum critical points determining an upper edge of the one-third plateau and saturation field, is gradually rounded upon raising temperature.   

\begin{figure}
\includegraphics[width=0.6\textwidth]{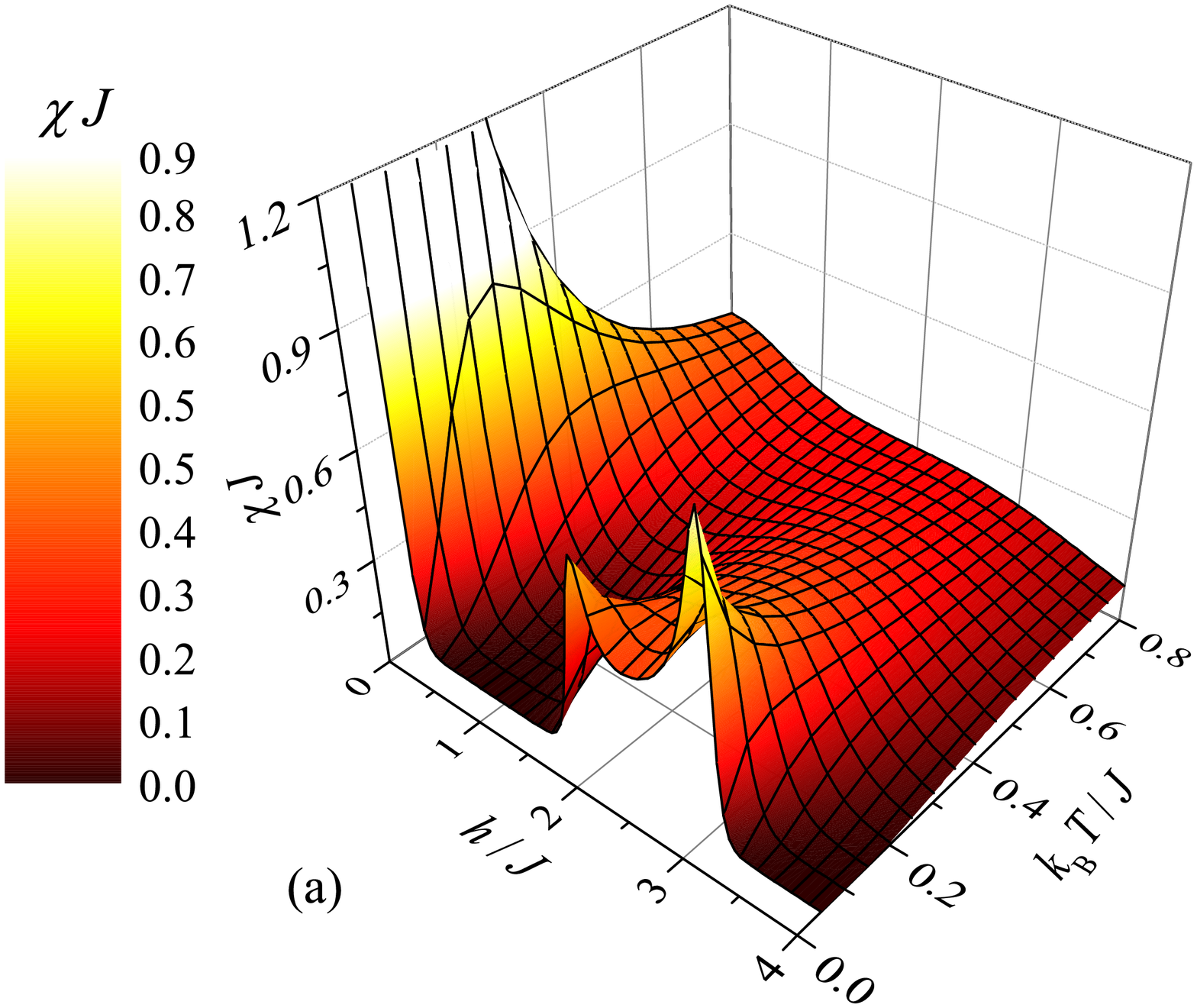}
\hspace{-1.7cm}
\includegraphics[width=0.6\textwidth]{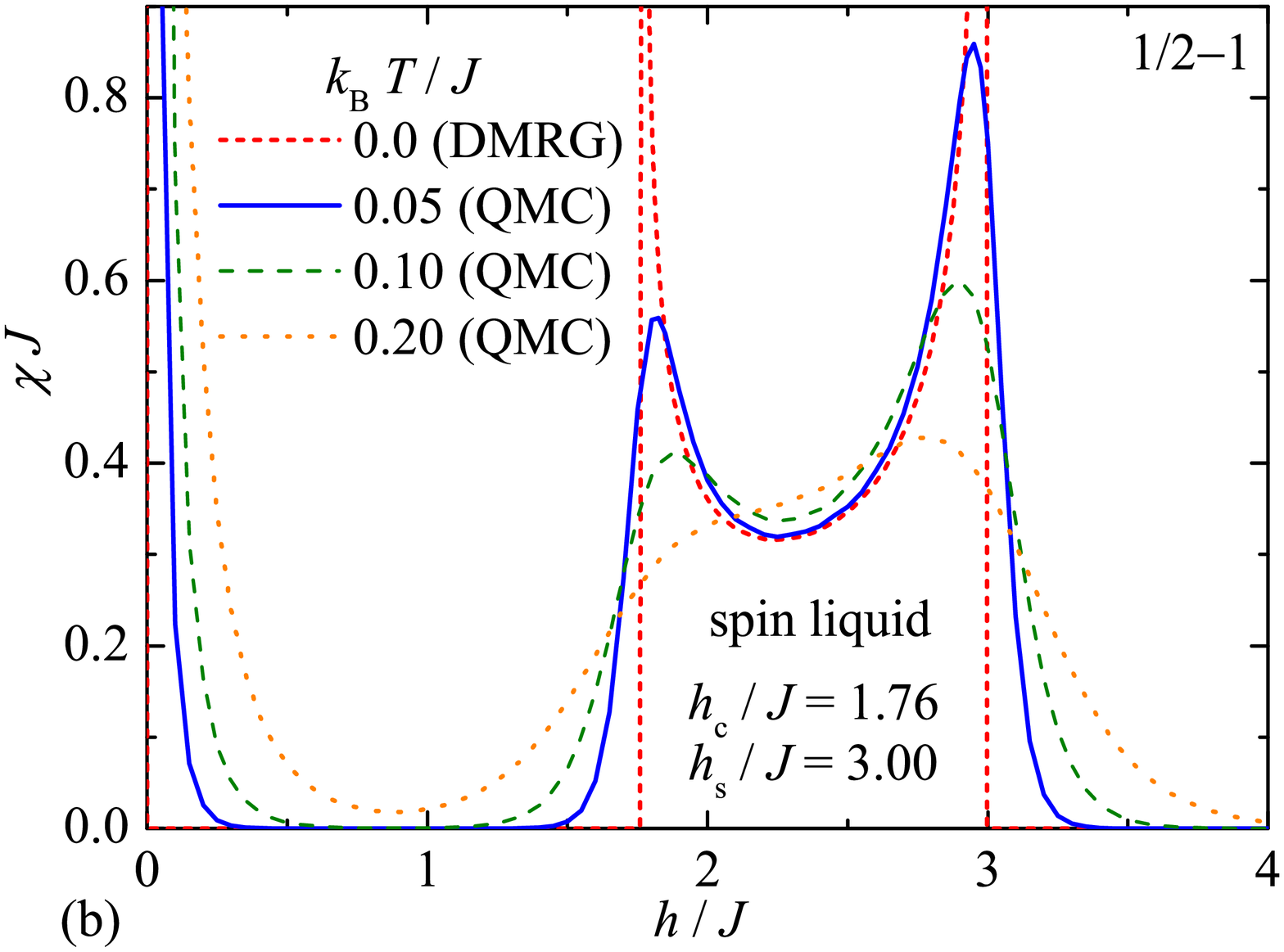}
\vspace*{-0.7cm}
\caption{The susceptibility of the ferrimagnetic mixed spin-(1/2,1) Heisenberg chain per unit cell: (a) 3D surface plot as a function of temperature and magnetic field constructed from QMC data; (b) Zero-temperature DMRG data versus QMC simulations at low temperatures.}
\label{fig2}       
\end{figure}

A stronger evidence of two quantum phase transitions of the ferrimagnetic mixed spin-(1/2,1)Heisenberg chain thus provides the susceptibility, which is plotted in Fig. \ref{fig2}. As a matter of fact, the susceptibility still displays at low enough temperatures two sharp peaks in a vicinity of the magnetic fields driving the investigated quantum spin system towards the quantum critical points in addition to a pronounced divergence observable at zero magnetic field. Of course, these local maxima are gradually suppressed and merge together upon increasing of temperature.

\begin{figure}
\includegraphics[width=0.6\textwidth]{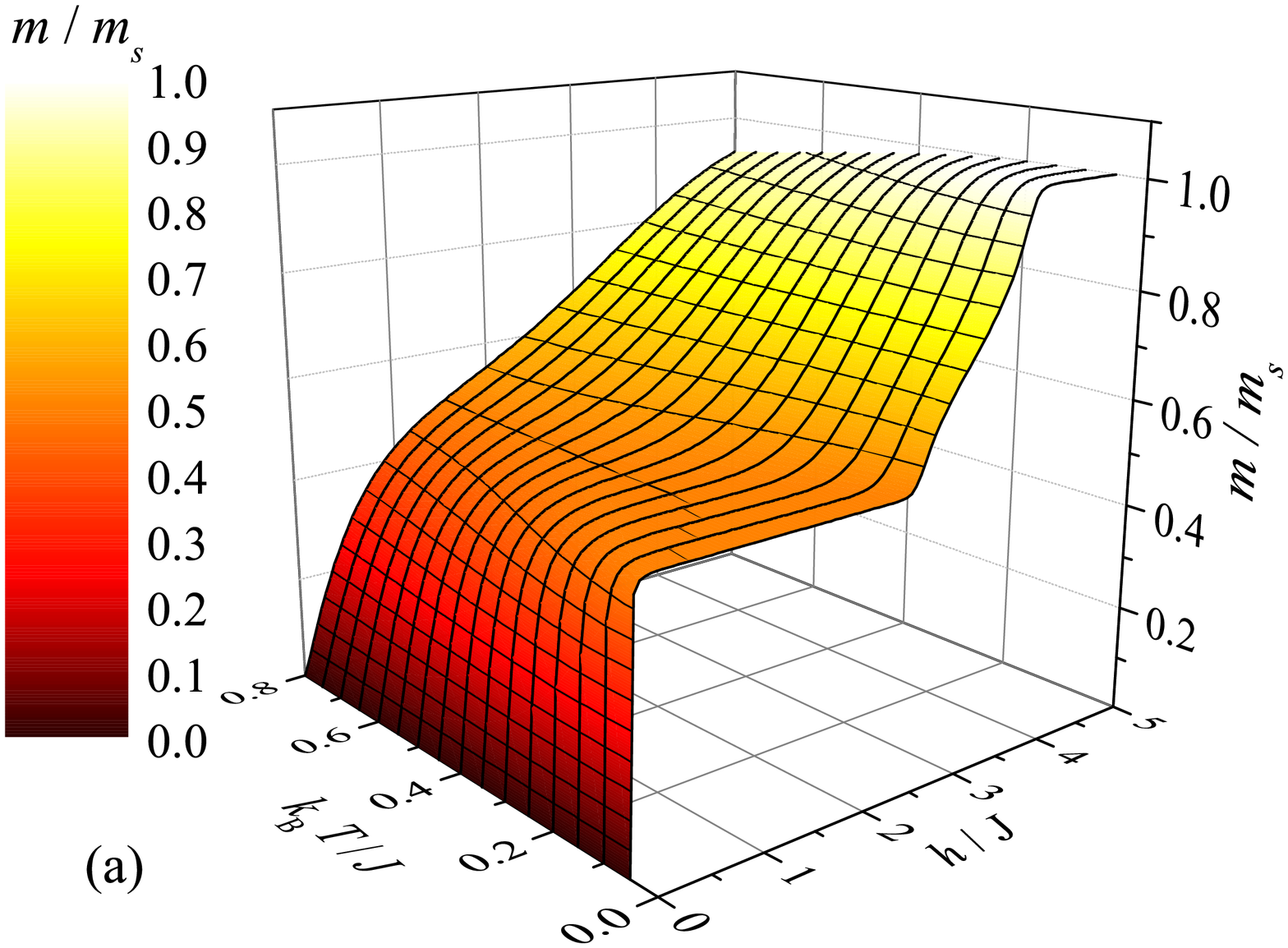}
\hspace{-1.7cm}
\includegraphics[width=0.6\textwidth]{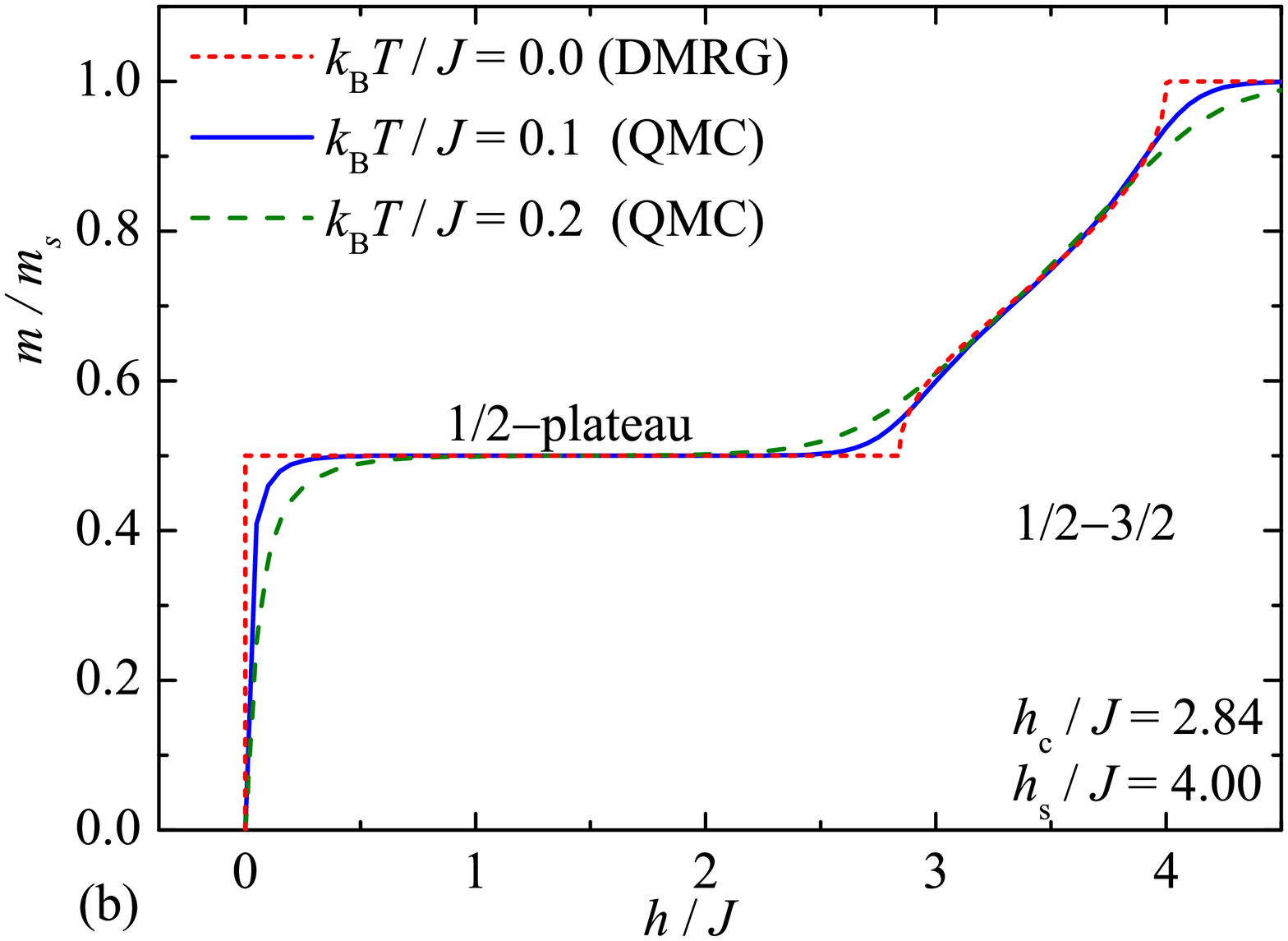}
\vspace*{-0.8cm}
\caption{The total magnetization of the ferrimagnetic mixed spin-(1/2,3/2) Heisenberg chain normalized with respect to its saturation value: (a) 3D surface plot as a function of temperature and magnetic field constructed from QMC data; (b) Zero-temperature DMRG data versus QMC magnetization curves at low enough temperatures.}
\label{fig3}       
\end{figure}

\begin{figure}
\includegraphics[width=0.6\textwidth]{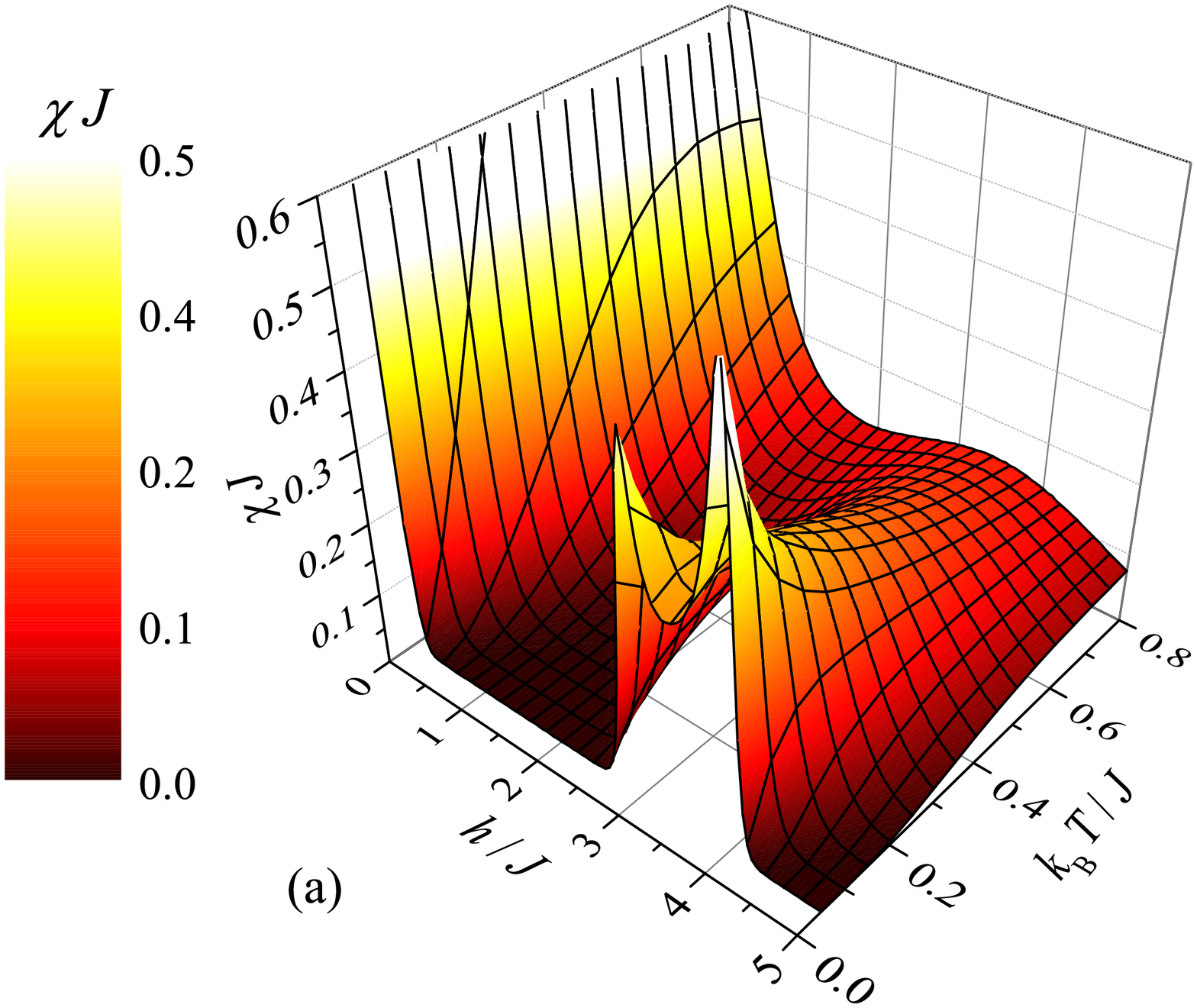}
\hspace{-1.7cm}
\includegraphics[width=0.6\textwidth]{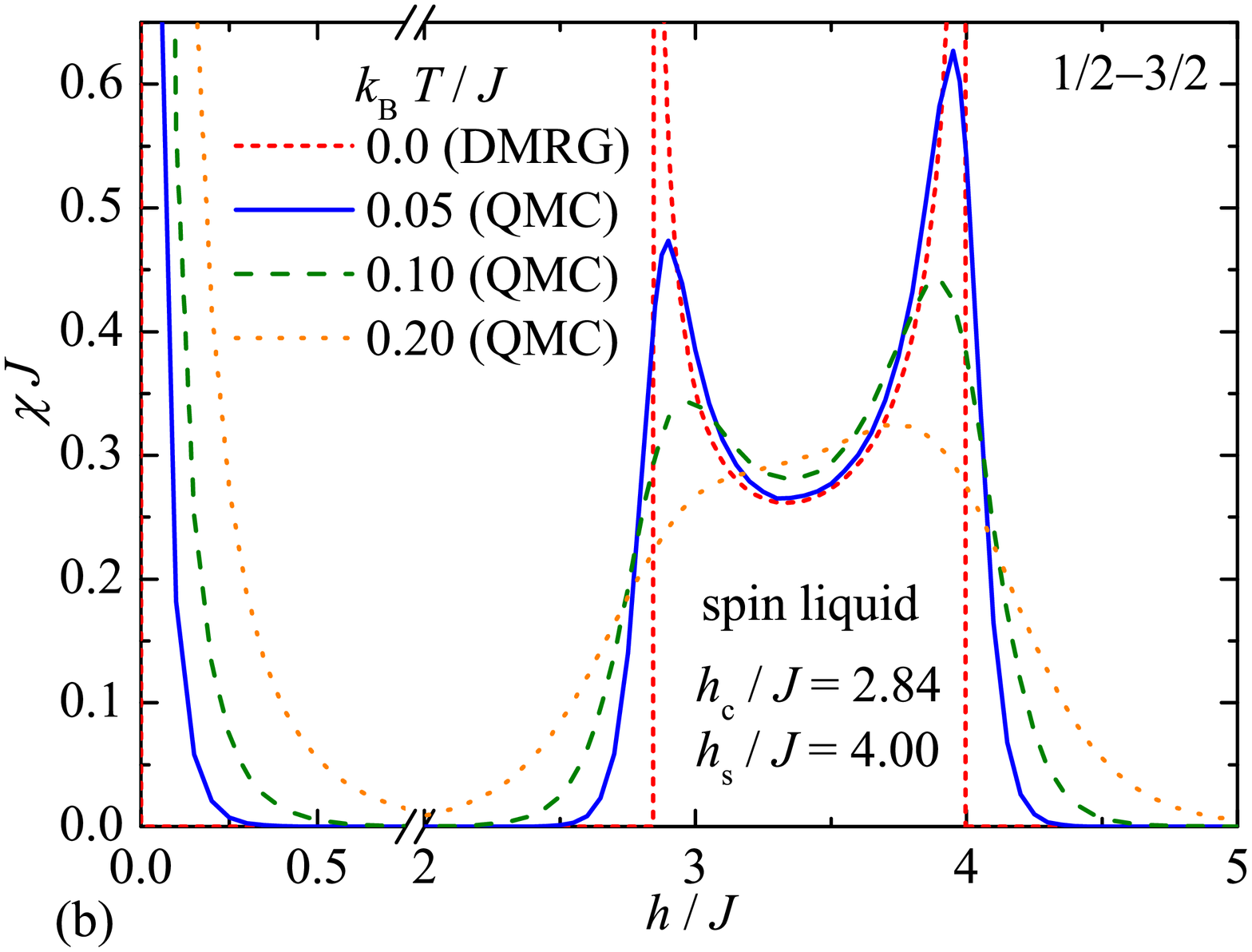}
\vspace*{-0.7cm}
\caption{The susceptibility of the ferrimagnetic mixed spin-(1/2,3/2) Heisenberg chain per unit cell: (a) 3D surface plot as a function of temperature and magnetic field constructed from QMC data; (b) Zero-temperature DMRG data versus QMC simulations at low enough temperatures.}
\label{fig4}       
\end{figure}

To illustrate a general validity of the aforedescribed conclusions, Figs. \ref{fig3} and \ref{fig4} show similar plots for the magnetization and susceptibility of another ferrimagnetic mixed spin-(1/2,3/2) Heisenberg chain. It is quite apparent that the same general trends can be reached as far as the temperature effect is concerned. There are only two quantitative differences. The first difference refers to a height of the intermediate Lieb-Mattis plateau, which is in the present case at one-half of the saturation magnetization. The second difference lies in absolute values of two field-driven quantum critical points, which determine the end of the intermediate one-half plateau and the beginning of the saturation magnetization, respectively.     

\begin{figure}
\includegraphics[width=0.55\textwidth]{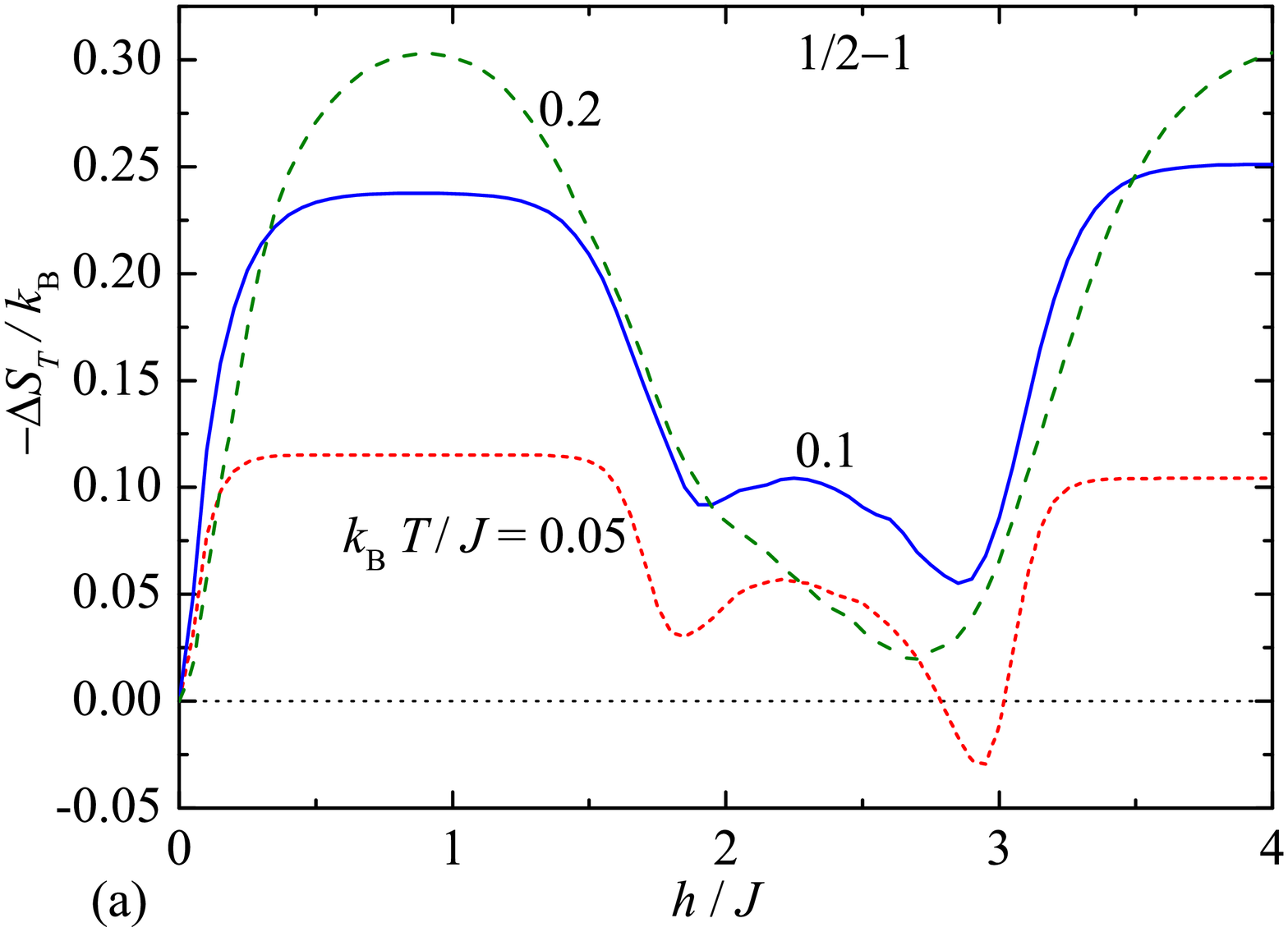}
\hspace{-1.0cm}
\includegraphics[width=0.55\textwidth]{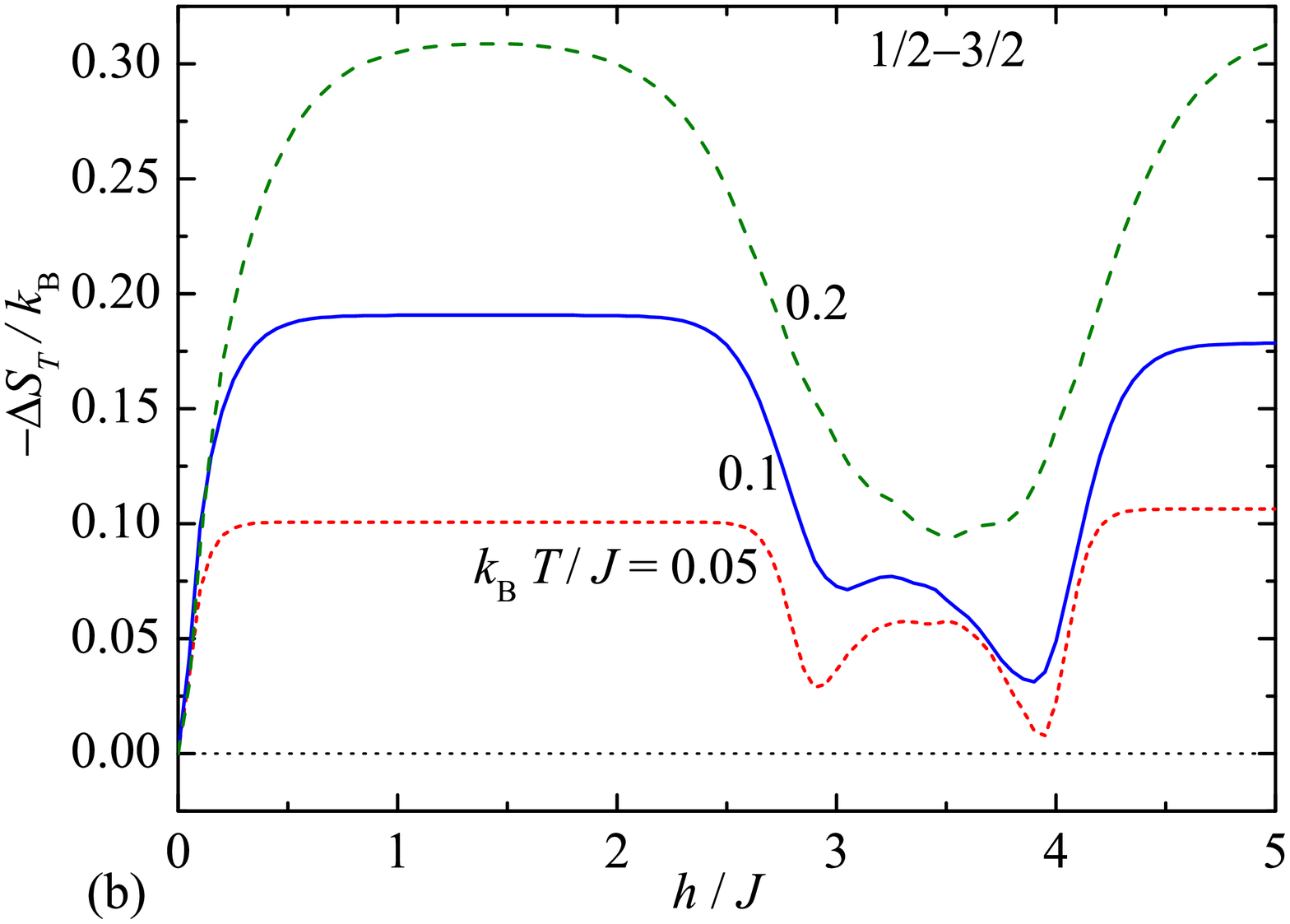}
\vspace*{-0.7cm}
\caption{The isothermal change of the entropy versus the magnetic field at three different temperatures for the ferrimagnetic mixed spin-(1/2,$S$) Heisenberg chain: (a) $S=1$; (b) $S=3/2$.}
\label{fig5}       
\end{figure}

Last but not least, let us examine the magnetic-field variations of the isothermal entropy change, which represents one of two basic magnetocaloric potentials. For this purpose, Fig.  \ref{fig5} illustrates typical dependences of the isothermal entropy change of two considered ferrimagnetic mixed spin-(1/2,$S$) Heisenberg chains as a function of the magnetic field at a few different temperatures. It is quite clear that the isothermal entropy change $-\Delta S_T$ exhibits a steep increase when the magnetic field is lifted from zero, which can be related to an abrupt field-induced increase of the magnetization from zero towards the ferrimagnetic Lieb-Mattis plateau $m/m_s= (2S-1)/(2S+1)$. The isothermal entropy change $-\Delta S_T$ then reaches a constant value within the field range, which is nearly equal to a width of the intermediate magnetization plateau at a given temperature. This interval of the magnetic fields determines range of applicability of the ferrimagnetic mixed spin-(1/2,$S$) Heisenberg chains for cooling purposes. Namely, the isothermal entropy change $-\Delta S_T$ consecutively acquires relatively deep minima near both field-driven quantum critical points, which are gradually lifted and smoothed upon increasing of temperature. The minima in the isothermal entropy change $-\Delta S_T$ can be thus regarded as another faithful indicators of the quantum critical points of the ferrimagnetic mixed spin-(1/2,$S$) Heisenberg chains.

\section{Conclusion}
\label{conc}
In the present work we have examined magnetic and thermodynamic properties of the ferrimagnetic mixed spin-(1/2,$S$) Heisenberg chains at finite temperatures using QMC simulations. In particular, we have focused our attention to a detailed examination of the temperature effect upon the magnetization process in a close vicinity of the field-driven quantum critical points. It has been evidenced that a singular square-root behavior of the magnetization, which accompanies both quantum phase transitions connected with a breakdown of the gapped Lieb-Mattis ferrimagnetic phase and the gapless Tomonaga-Luttinger spin-liquid phase, undergoes a gradual rounding upon increasing of temperature. The magnetization curve at non-zero temperatures is thus almost without any clear signature of the quantum critical points unlike other thermodynamic response functions.

The susceptibility and isothermal entropy changes contrarily display close to quantum phase transitions relatively sharp maxima and minima, respectively, which are gradually suppressed and smoothed upon increasing of temperature. The respective maximum in the susceptibility and minimum in the isothermal entropy change can be accordingly regarded as a faithful indicator of the field-driven quantum critical point of the ferrimagnetic mixed spin-(1/2,$S$) Heisenberg chains. Although the present study was restricted just to the mixed-spin Heisenberg chains with the quantum spin numbers $S=1$ and 3/2, the foregoing studies \cite{kur98,yam99,sak99,iva00,hon00,yam00,sak02,ten11,iva16,str16} have already proven largely universal behavior of the ferrimagnetic mixed spin-(1/2,$S$) Heisenberg chains also for other spin values for which qualitatively the same behavior should be expected. 

It has been also demonstrated that the most efficient cooling through the adiabatic demagnetization of the ferrimagnetic mixed spin-(1/2,$S$) Heisenberg chains can be achieved in a range of the magnetic fields from zero up to nearly a half of the intermediate magnetization plateau, while the isothermal entropy change implies for larger magnetic fields a less efficient cooling as it rapidly drops down near both quantum critical points. It is our hope that theoretical results of the present work are of obvious relevance for a series of bimetallic coordination compounds MM'(pba)(H$_2$O)$_3$ $\cdot$ 2H$_2$O \cite{kah87} and  MM'(EDTA) $\cdot$ 6H$_2$O \cite{dri85} (M,M' = Cu, Ni, Co, Mn), which represent experimental realizations of the ferrimagnetic mixed-spin Heisenberg chains. High-field magnetization and magnetocaloric measurements on these or related series of bimetallic complexes are however needed for experimental testing of the present theoretical predictions.

\end{document}